\documentclass[aps,prl,twocolumn,superscriptaddress,showpacs]{revtex4}

\usepackage{epsfig}
\usepackage{amsmath}
\usepackage{amssymb}

\begin{document}

\title{Quantum Hall States at $\nu=\frac{2}{k+2}$}

\author{Waheb Bishara}
\affiliation{Department of Physics, California Institute of Technology, MC 114-36, Pasadena, California 91125, USA}
\author{Gregory A. Fiete}
\affiliation{Department of Physics, California Institute of Technology, MC 114-36, Pasadena, California 91125, USA}
\author{Chetan Nayak}
\affiliation{Microsoft Research, Station Q, CNSI Building, University of California, Santa Barbara, California 93106, USA}
\affiliation{Department of Physics, University of California, Santa Barbara, California, 93106, USA}

\date{\today}

\begin{abstract}
We study the $\nu=\frac{2}{k+2}$ quantum Hall states which are
particle-hole conjugates of the $\nu=\frac{k}{k+2}$ Read-Rezayi states.
We find that equilibration between the different modes at the edge
of such a state leads to an emergent SU(2)$_k$ algebra in the counter-propagating neutral sector.
Heat flow along the edges of these states will be in the opposite direction of
charge flow. In the $k=3$ case, which may be relevant to $\nu=2+\frac{2}{5}$,
the thermal Hall conductance and the exponents associated
with quasiparticle and electron tunneling distinguish this state
from competing states such as the hierarchy/Jain state.
\end{abstract}



\maketitle

The most robust state in the
second Landau level (SLL) is the $\nu=5/2$ state \cite{Willet:prl87,Pan:prl99,EisentsteinPRL02}.
As a result of its even-denominator, it cannot belong to the
usual hierarchy/`composite fermion' sequence of Abelian states \cite{Laughlin83,Haldane83,Halperin84,Jain89,Read90,Wen92}
which seems to explain all of the observed states in the
lowest Landau level (LLL).
The leading candidate theories of the $\nu=5/2$ state are non-Abelian:
the Pfaffian state \cite{Moore:npb91,Greiter92,Nayak96c}
and its particle-hole conjugate, the anti-Pfaffian state \cite{Levin07,Lee07}.
Thus, one may wonder whether the other fractions observed
in the second Landau level, such as $\nu=7/3$, $12/5$, $8/3$ and $14/5$
\cite{Pan:prl99,Xia04,Choi08}, are also non-Abelian even though they occur at
odd-denominator filling fractions. The state at $\nu=12/5=2+\frac{2}{5}$
has been the subject of particular interest because its filling fraction
is the particle-hole conjugate\footnote{We note that the $k=2,3$ RR states in the SLL
appear to be weaker than the corresponding $\overline{\rm RR}$ states,
in contrast with the LLL, where the Jain states are stronger than their particle-hole conjugates.} of that of the $k=3$ Read-Rezayi state
\cite{Read99}.
This is an exciting possibility because this state is capable of
supporting universal topological quantum computation
\cite{Freedman02a,Nayak08}. Alternatively, a state at the lowest level of
a non-Abelian hierarchy built on a $\nu=5/2$ Pfaffian
state also occurs at $\nu=12/5$ \cite{BondersonHierarchy}.
Finally, the $\nu=12/5$ state may simply be the transposition to
the second Landau level of the Abelian state which is believed
to occur at $\nu=2/5$.

In this paper, we analyze the particle-hole conjugates of the
general level-$k$ Read-Rezayi states, which we call the level-$k$
$\overline{\text{RR}}$ states. These states possess multiple
gapless modes of edge excitations, which are of particular
interest for charge and heat transport.
We formulate the low-energy effective field theories of
the edges of the level-$k$ $\overline{\text{RR}}$ states
and show that an SU(2)$_k$ Kac-Moody symmetry emerges
when the different edge modes equilibrate.
One notable feature is that the thermal Hall conductance due to this state
$\kappa_{xy}=-\left(\frac{2k-2}{k+2}\right){\pi^2 k_B^2 T}/{3h}$,
is opposite in sign to the electrical Hall
conductance, $\sigma_{xy}=\frac{2}{k+2}\frac{e^2}{h}$.
We then focus on the $k=3$ $\overline{\text{RR}}$ state
and compare it to other possible $\nu=2+2/5$ states.
We show that charge transport through a quantum point contact
and thermal transport can distinguish this state from its competitors.

The action of the edge of the level $k$-RR state \cite{Read99}
at filling fraction $\nu=\frac{k}{k+2}$ is composed of charged and neutral sectors. The charged sector is described by
a chiral bosonic field propagating with velocity $v_c$.
The neutral sector is a chiral $\mathbb{Z}_k$ parafermionic theory \cite{Fateev85}
propagating with velocity $v_n$. The $\mathbb{Z}_k$ parafermion
theory is an SU(2)$_k$/U(1) coset with central charge
$c=\frac{2k-2}{k+2}$ which
can be represented by an SU(2)$_k$ chiral WZW model in which
the U(1) subgroup has been gauged \cite{Karabali90} (note that the gauge field
is not minimally coupled \cite{Witten84}). Thus, we can write:
\begin{multline}
\label{eq:S-RR}
S = \frac{1}{4\pi \nu} \int dx d\tau\, {\partial_x} \varphi\overline{\partial} \varphi
\: + \: S_{{\rm WZW},k}\\
 + \frac{k}{4\pi} \int dx d\tau \, \text{tr}
\Bigl({A_x}\overline{\partial}g \cdot g^{-1}
- \overline{A}g^{-1}{\partial_x}g
+ {A_x}g  \overline{A}g^{-1}
- {A_x} \overline{A}
\Bigr),
\end{multline}
where $\overline{\partial}\equiv i{\partial_\tau} + {v_c}{\partial_x}$
and $\overline{A}\equiv A_{\tau}-i{v_n}{A_x}$. The neutral sector is
the sum of the second and third terms which we will call
$S^{}_{\mathbb{Z}_k} =\int {\cal L}^{}_{\mathbb{Z}_k}$.
The second term, the WZW action, is given by:
\begin{multline}
\label{eqn:WZW}
S_{{\rm WZW},k} = \frac{k}{16\pi}\int d\tau dx\,
\text{tr}\left( {\partial_x} {g^{-1}}
\overline{\partial} g\right)
\\ -\:
i\frac{k}{24\pi}\int dx d\tau dr\,  \epsilon^{\mu\nu\lambda}
\text{tr}\left( {\partial_\mu} g\,{g^{-1}}
{\partial_\nu} g\,{g^{-1}}\,{\partial_\lambda}g\,{g^{-1}}\right).
\end{multline}
The field $g$ takes values in SU(2).
The second integral is over any three-dimensional manifold $M$
which is bounded by the two-dimensional spacetime of the edge $\partial M$.
The value of this integral depends only on the values of the field $g$
at the boundary $\partial M$.
As a result of the gauging (\ref{eq:S-RR}), the primary
fields $\Phi_{j,m}$ of this model are essentially the spin-$j$ primary fields
of the WZW model (\ref{eqn:WZW}) dressed by charge-$m$
Wilson lines of the U(1) gauge field; as a result of the latter,
they are invariant under the U(1) subset of the gauge group.
The $k(k+1)/2$ primary fields $\Phi_{j,m}$ are, consequently,
indexed by half-integers $j,m$
satisfying $0\leq j\leq k/2$, $m\in(-j,-j+1,\ldots,j)$ with the identifications
$(j,m)\cong(\frac{k}{2}-j,m+\frac{k}{2})$, $(j,m)\cong(j,m+k)$.
The field $\Phi_{j,m}$ has dimension $\Delta_{j,m}=\frac{j(j+1)}{k+2} - \frac{m^2}{k}$.
Of particular importance is the parafermion field
${\psi_1}\equiv \Phi_{\frac{k}{2},-\frac{k}{2}+1}$
of dimension $\Delta=1-\frac{1}{k}$. For $k=1$, the theory is trivial;
the $k=1$ RR state is simply the $\nu=1/3$ Laughlin state which
has no neutral sector. The $k=2$ RR state is the Pfaffian state; in the special
case $k=2$, the SU(2)$_2$/U(1) coset can be alternately
represented as a Majorana fermion.
The three primary fields are then $\Phi_{0,0}=1$, $\Phi_{1/2,1/2}=\sigma$,
$\Phi_{1,0}=\psi$.

In the RR state, the electron creation operator is a charge-1
fermionic operator, $\Psi^{\dagger}_{e}={\psi_1} e^{i\frac{k+2}{k}\phi}$,
where $\psi_1$ is the $\mathbb{Z}_k$ parafermion field described
above (simply the Majorana fermion in the $k=2$ case).
With the bosonic field $\phi$ normalized as in (\ref{eq:S-RR}),
the scaling dimension of $e^{i\alpha\phi}$ is $\nu \frac{\alpha^2}{2}$.
Consequently, the electron operator has scaling dimension $\frac{3}{2}$.
The neutral sector does not enter the charge current,
$J=\frac{1}{2\pi}\partial\varphi$, so the
level-$k$ RR state has a quantized Hall conductance
$\sigma_{xy}=\frac{k}{k+2}\,\frac{e^2}{h}$. If this fractional quantum
Hall state occurs in the second Landau level and the lowest Landau
level (of both spins) is filled and inert, then $\sigma_{xy}=\left(2+\frac{k}{k+2}\right)
\frac{e^2}{h}$. The energy momentum tensor is the sum of the
two energy momentum tensors, $T={T_c}+T_{\mathbb{Z}_k}$.
Consequently, the thermal Hall conductivity is proportional to the sum of the
two central charges \cite{ReadGreen2000}: $\kappa_{xy}=\frac{3k}{k+2}\,\frac{\pi^2 k_B^2}{3h}T$.
If this this fractional quantum Hall state occurs in
the second Landau level, then
$\kappa_{xy}=\left(2+\frac{3k}{k+2}\right)\frac{\pi^2 k_B^2}{3h}T$.

To find the edge structure of the level $k$ anti-RR state ($\overline{RR}$),
we generalize the analysis done for $k=1$ in Ref. \cite{Kane94} and for
$k=2$ in Refs. \cite{Levin07,Lee07}. Ignoring filled Landau
levels (if any), we perform a particle-hole transformation
of the partially filled Landau level (the second Landau level in the case
of $\nu=12/5$). The edge between the level-$k$ $\overline{\text{RR}}$
state ($\nu=1-\frac{k}{k+2}=\frac{2}{k+2}$) and the vacuum ($\nu=0$) is mapped
to the edge between the level $k$-RR state ($\nu=\frac{k}{k+2}$) and a $\nu=1$ state.
Hence, the theory of this edge is described by a level $k$-RR edge theory
and a counter propagating bosonic charge mode which is the edge theory
of the $\nu=1$ state. The low-energy effective Lagrangian is:
\begin{multline}
\label{eqn:anti-RR}
{\cal L}_{\overline{RR}} =
\frac{1}{4\pi} \partial_x \phi_1(i\partial_\tau + v_1 \partial_x) \phi_1\\
+\left(\frac{k+2}{k}\right)\frac{1}{4\pi} \partial_x \phi_2(-i\partial_\tau + v_2 \partial_x) \phi_2
\: + \: {\cal L}^{}_{\mathbb{Z}_k} \\
- \frac{2}{4\pi} v_{12} \partial_x \phi_1 \partial_x \phi_2 + \xi(x)\psi_1 e^{i\frac{k+2}{k}\phi_2}e^{-i\phi_1} +h.c.,
\end{multline}
where $\phi_1$ is the $\nu=1$ edge charge mode and $v_{12}>0$
is a repulsive density-density interaction along the edge.
The final term is inter-mode electron tunneling which tunnels electrons
from the outer $\nu=1$ edge to the inner edge with a random coefficient $\xi$ which,
for simplicity, we take to be of Gaussian white noise form:
$\langle \xi(x) \xi^*(x')\rangle=W\delta(x-x')$.
In the absence of inter-mode tunneling, this theory will
not realize a universal value of the two-terminal conductance.
The tunneling term allows the counter-propagating modes to
equilibrate and achieve a universal two-terminal conductance,
as is the case for the $\nu=2/3$ quantum Hall state \cite{Kane94}.

For $v_{12}=0$, the inter-mode electron tunneling term is irrelevant:
${dW}/{d\ell}=-W$, as may be seen by using the replica trick to integrate
out $\xi$. However, for $v_{12}$ sufficiently large, $W$ becomes relevant.
To see this, we introduce a new set of fields defined by
\begin{equation}
\phi_\rho=\phi_1-\phi_2 \, , \;\;\;
 \phi_\sigma = \phi_1 - {\phi_2} (k+2) /k,
\end{equation}
corresponding to charged and neutral bosonic modes, respectively. In these variables,
the Lagrangian takes the form
${\cal L}_{\overline{RR}}={\cal L}_\rho + {\cal L}_{\sigma} + {\cal L}_{\rm tun}+ {\cal L}_{\rho\sigma}$,
with:
\begin{align}
\mathcal{L}_{\rho}&= \frac{1}{4\pi}\left(\frac{k+2}{2}\right) \partial_x \phi_\rho(i\partial_\tau +
v_\rho \partial_x) \phi_\rho, \cr
\mathcal{L}_{\sigma}&= \frac{1}{4\pi}\frac{k}{2}\, \partial_x \phi_\sigma(-i\partial_\tau
 + v_\sigma \partial_x) \phi_\sigma +{\cal L}_{\mathbb{Z}_k}(v_n) ,\cr
\mathcal{L}_{\rho\sigma}&= 2 v_{\rho\sigma} \partial_x \phi_\sigma \partial_x \phi_\rho,\cr
\mathcal{L}_{\rm tun}&= \xi(x)\psi_1e^{i\phi_\sigma}+\xi^*(x) \psi_1^{\dagger}e^{-i\phi_\sigma},
\end{align}
and $v_\sigma$, $v_\rho$, $v_{\rho\sigma}$ are functions of $v_1$,$v_2$ and $v_{12}$,
e.g. $4\pi v_{\rho\sigma}= {(k/2)^2}{v_1}+{(k+2/2)^2}{v_2}-({k(k+2)/4+(k/2)^2})v_{12}$.
If $v_{\rho\sigma}=0$, then the electron tunneling operator has scaling dimension
$[\psi_1 e^{i\phi_\sigma}]=1$ and the inter-mode electron tunneling term is {\it relevant}:
${dW}/{d\ell}=W$.

We now show that when the disorder is a relevant perturbation,
the edge theory flows to a new fixed point described by a
freely-propagating charged boson (responsible for the universal quantized Hall conductance)
and a backward propagating neutral sector that possesses an $SU(2)$ symmetry.
We will argue that due to the disordered tunneling the neutral modes will equilibrate
and propagate at common average velocity $\bar{v}$ and show that the velocity mismatch
and the mixing term $\mathcal{L}_{\rho\sigma}$ are irrelevant.
An $SU(2)$ symmetry will thus emerge in the neutral sector.
Note that for $k=2$ this reduces to the result obtained for the anti-Pfaffian \cite{Levin07,Lee07}.
Let us write the neutral sector action $\mathcal{L}_\sigma$ as
$\mathcal{L}_{SU(2)_k}+\mathcal{L}_{\delta v}$, with:
\begin{eqnarray}
\mathcal{L}_{SU(2)_k}&=&\frac{1}{4\pi}\frac{k}{2} \partial_x \phi_\sigma(-i\partial_\tau +
\bar{v} \partial_x) \phi_\sigma +{\cal L}_{\mathbb{Z}_k}(\bar{v}),\\
\mathcal{L}_{\delta v}&=&  \left(\mathcal{L}_{\mathbb{Z}_k}(v_n) - \mathcal{L}_{\mathbb{Z}_k}(\bar{v})\right) +\frac{1}{4\pi}\frac{k}{2}(v_\sigma-\bar{v})(\partial_x \phi_\sigma)^2. \nonumber
\end{eqnarray}
The Lagrangian $\mathcal{L}_{SU(2)_k}$ is, in fact, equivalent to (the opposite
chirality version of) the chiral WZW action (\ref{eqn:WZW}):
the chiral boson $\phi_\sigma$ restores the U(1) which
was gauged out in (\ref{eq:S-RR}). A simple way to see this is to note that
the currents:
\begin{equation}
\label{ParafCurrents}
J^+ = \sqrt{k}\psi_1 e^{i\phi_\sigma},\;\;
J^{-} = \sqrt{k}\psi_1^{\dagger} e^{-i\phi_\sigma}, \;\;J^z=\frac{k}{2}\partial_x \phi_\sigma,
\end{equation}
obey the same SU(2)$_k$ Kac-Moody commutation relations as the WZW currents:
\begin{equation}
J^a=-\frac{ik}{2\pi}tr\left(T^a g^{-1} (i\partial_\tau-\bar{v}\partial_x)g\right),
\end{equation}
where $T^a$, $a=x,y,z$ are $SU(2)$ generators and
$J^\pm = {J^a} \pm i {J^y}$.

We notice that the tunneling term $\mathcal{L}_{\rm tun}$ can be written in terms of the currents:
\begin{equation}
\mathcal{L}_{\rm tun}=\xi(x) J^+ +\xi^*(x) J^-.
\end{equation}
It is convenient to use the WZW representation since the tunneling term
can be eliminated from the action by the gauge transformation
$g\rightarrow gU$ with $U=Pe^{\frac{i}{\bar{v}}\int^x dx' \vec{\xi}(x')\cdot \vec{T}}$,
where $P$ denotes path ordering and
$\vec{\xi}(x')=(2\text{Re}(\xi(x')),-2\text{Im}(\xi(x')),0)$.
Under this gauge transformation
$\mathcal{L}_{SU(2)_k} \rightarrow \mathcal{L}_{SU(2)_k} - \vec{\xi}\cdot\vec{J}$,
thus gauging away the tunneling term $\mathcal{L}_{\rm tun}$.

We now turn to the effect of this gauge transformation
on the velocity anisotropy terms. The velocity terms in $\mathcal{L}_{\sigma}$
can be written in the form:
\begin{equation}
v_a {\rm tr} \left(S_a \partial_x g^{-1} \partial_x g \right),
\end{equation}
where $S_a$ is a matrix satisfying $\rm{tr}(S_aT_bT_c)=\delta_{ab}\delta_{ac}$
and $v_a$, $a=x,y,z$ can be expressed in terms of $v_\sigma$, $v_n$.
Let us separate the traceless part $M$ of the matrix $v_a S_a$:
$M = v_a S_a - {\rm tr} ( v_a S_a )\times \mathbb{I}/3$.
Then $\mathcal{L}_{\delta v}$ takes the form
$\mathcal{L}_{\delta v} =  {\rm tr} (M \partial_x g^{-1} \partial_x g)$
Under the gauge transformation $g\rightarrow gU$,
$\mathcal{L}_{\delta v}  \rightarrow tr (M' \partial_x g^{-1} \partial_x g)$,
where $M'=UMU^{\dagger}$ is random since the gauge transformation
is a function of $\xi(x)$.
The renormalization group flow of the mean square average of $M'$,
$W_{M'}$, is $dW_{M'}/dl=(3-2\Delta)W_{M'}$
\cite{GiamarchiSchultz}, where $\Delta$ is the scaling dimension of the term to which $M'$ couples. In this case, $M'$ couples to $\partial_x g^{-1} \partial_x g \propto J^2$
which has scaling dimension $\Delta=2$ (i.e. $M'$ is a velocity).
Hence $W_{M'}$ and the velocity anisotropy are irrelevant.
The part of the velocity term which is invariant under the gauge transformation
is the average velocity $\bar{v}={\rm tr} ( v_a S_a )/3$.

The mixing term, $\mathcal{L}_{\rho\sigma}$ is irrelevant. It
can be written as $\mathcal{L}_{\rho\sigma}=
2v_{\rho\sigma} (\frac{2}{k}J^z)\cdot \partial_x \phi_\rho$;
under the gauge transformation $g\rightarrow gU$ the current $J^z$
gets rotated with a random coefficient. Consequently, deviations from $v_{\rho\sigma}=0$
are irrelevant, much like the velocity anisotropy term above.

Thus, we have found that at the fixed point where the edge modes equilibrate
due to random electron tunneling, the edge theory of the anti-RR state is described
by a single bosonic charge mode, $\mathcal{L}_\rho$, and an
$SU(2)_k$ neutral sector, $\mathcal{L}_{SU(2)_k}$, moving in the opposite direction.
The electron operator of the $\nu=1$ edge in the unequilibrated theory with $\xi=0$ in
(\ref{eqn:anti-RR}) is $e^{i\phi_1}$, which can be rewritten in the form
$e^{i\phi_1}=e^{-i\frac{k}{2}\phi_\sigma}\,e^{i\frac{k+2}{2}\phi_\rho}$. As a result
of equilibration, the dimension of this operator changes, from ${\Delta_e}=1/2$ to
${\Delta_e}=(k+1)/2$. (The conformal spin, the difference between the
right and left scaling dimensions, remains $1/2$, however.) Noting that
$e^{i\phi_1}$ can be rewritten as $\chi_{j=k/2}^{m=-k/2} \,e^{i\frac{k+2}{2}\phi_\rho}$, we see that
this operator is the lowest $J^z$ eigenvalue, $m$, of a multiplet
$\chi_{j=k/2}^{m} \,e^{i\frac{k+2}{2}\phi_\rho}$ with $m=-k/2,-k/2+1,\ldots,k/2$.
The other electron creation operators in this SU(2) mutiplet are obtained by acting multiple times on $e^{i\phi_1}$
with $J^+ = \psi_1 e^{i\phi_\sigma}$; thus, they create an electron in the original
$\nu=1$ edge and transfer multiple electrons from the RR edge to the $\nu=1$ edge.
As a result of equilibration, all $k+1$ of these operators have the same scaling dimension.
When electrons tunnel between two level-$k$ $\overline{RR}$ droplets,
the tunneling conductance $G\sim T^{2k}$ and, for finite $V>T$, $I_{\rm tun}\sim V^{2k+1}$.

Quasiparticle operators can be obtained by the requirement that they are
local with respect to these electron operators. The allowed quasiparticle operators
(modulo the creation or annihilation of an electron) and their scaling dimensions are:
\begin{equation}
\Phi_{\rm qp}^{j,N} = {\chi_j} \, e^{i(j+N)\phi_\rho}.
\end{equation}
The $J^z$ eigenvalue is suppressed here; there is an SU(2) multiplet of each of
these operators all of which belong to the same quasiparticle species because
they have the same topological properties. $\Phi_{\rm qp}^{j,N}$ has right scaling
dimension ${(j+N)^2}/(k+2)$ and left scaling dimension ${j(j+1)}/(k+2)$ and, therefore,
total scaling dimension $\left[{(j+N)^2}+{j(j+1)}\right]/(k+2)$ and
topological spin $\left[{(j+N)^2}-{j(j+1)}\right]/(k+2)$.
For $k$ even, $N=0,1,\ldots,\frac{k}{2}$. For $k$ odd, $N=0,1,\ldots,\frac{k+1}{2}$
for integer $j$ and $N=0,1,\ldots,\frac{k-1}{2}$ for half-integer $j$. Therefore,
there are $(k+1)(k+2)/2$ different quasiparticle species. This is also the ground
state degeneracy of the $\overline{RR}$ theory on the torus (which is 10 in the case of
the $\nu=12/5$ state). The corresponding RR state has the same degeneracy.
The minimal dimension of a quasiparticle operator is
$[\Phi_{\rm qp}^{1/2,0}]=[\Phi_{\rm qp}^{0,1}]=\frac{1}{k+2}$.
Consequently, when quasiparticles tunnel between the edges
at a point contact, $R_{xx}\sim T^{-2k/(k+2)}$ and, at finite $V>T$,
$I_{\rm tun}\sim V^{(2-k)/(2+k)}$.

The thermal Hall conductivity of the anti-RR state is determined by
the central charge of the edge theory \cite{ReadGreen2000}. Ignoring the filled Landau levels,
the central charge of the
bosonic charge sector is $c=1$ and the central charge of the
$SU(2)_k$ theory is $\overline{c}=3k/(k+2)$. The thermal Hall
conductivity of the anti-RR state is then:
\begin{equation}
\kappa_{xy}^{\overline{RR}}=\left(1-\frac{3k}{k+2}\right)\frac{\pi^2 k_B^2}{3h}T.
\end{equation}
Thus, the conductivity due to the partially filled second Landau level
is negative for all $k$. Focusing on the $\nu=2/5$ anti-RR state (k=3)
its thermal Hall conductivity is $-\frac{4}{5}$ (in units of $\frac{\pi^2 k_B^2}{3h}T$),
while the Abelian hierarchy state at $\nu=2/5$ has a positive thermal Hall conductance of $+2$,
and the $\nu=2/5$ non-Abelian hierarchy
state of Ref.[\onlinecite{BondersonHierarchy}],
built on the $\nu=1/2$ Pfaffian state, would have a thermal Hall
conductance of $+\frac{1}{2}$. We note that the construction of
Ref.[\onlinecite{BondersonHierarchy}] can also
produce a $\nu=2/5$ state built on the anti-Pfaffian state, with
thermal Hall conductance $-\frac{3}{2}$.
These thermal conductivities are achieved at length scales longer that the
equilibration length of the edges. In the case of the $\nu=12/5$ state,
the filled lower Landau level gives an additional contribution of $+2$,
which would make all of the thermal conductivities positive, though
differing in magnitude. Therefore, in order to distinguish the non-Abelian
$\nu=12/5$ states from the Abelian one through the signs
of their thermal Hall conductivities, it would be necessary to measure
the thermal conductivity along an edge between
$\nu=2$ and $\nu=2+\frac{2}{5}$, which would only have a contribution
from the partially-filled Landau level.
On shorter length scales, the different modes on the edge do not equilibrate,
in which case both the anti-RR state and the non-Abelian hierarchy state
will have heat flow both upstream and downstream while the
Abelian state will have purely chiral heat transport. In this case,
the filled Landau levels simply give an additional contribution to the
downstream heat transport.

The difference between the various proposed $\nu=12/5$ states
would also be evident from the transport through a point contact.
As a result of weak quasiparticle tunneling from one edge to the other,
there is a non-zero longitudinal resistance $R_{xx} \sim T^{4\Delta_{\rm qp}-2}$.
At finite voltage $V>T$, we instead have
$I_{\rm tun}\sim V^{4\Delta_{\rm qp}-1}$.
In the Abelian hierarchy $\nu=2/5$ state,
the most relevant tunneling operator is that of the charge $\frac{2}{5}e$
quasiparticle with $\Delta_{\rm qp}=\frac{1}{5}$ \cite{Wen92,Wen95},
leading to $R_{xx} \sim T^{-6/5}$.
In the non-Abelian hierarchy state of Ref. [\onlinecite{BondersonHierarchy}],
the most relevant tunneling operator is that of
charge $\frac{1}{5}e$ quasiparticles with dimension
$\Delta_{\rm qp}=\frac{9}{80}$, leading to $R_{xx} \sim T^{-31/20}$.
Its sister state, built on the anti-Pfaffian, rather than the Pfaffian
has $\Delta_{\rm qp}=\frac{19}{80}$, hence $R_{xx} \sim T^{-21/20}$.
Finally, in the $k=3$ $\overline{\rm RR}$ state,
the operator $\Phi_{1/5}=\Phi_{(\frac{1}{2})}\,e^{i\frac{1}{2}\phi_\rho}$
carries charge $\frac{1}{5}e$ and has scaling dimension
$\Delta_{\rm qp}=\frac{1}{5}$,
while the operator $\Phi_{2/5}=e^{i\phi_\rho}$ carries charge
$\frac{2}{5}e$ and has the same scaling dimension. Therefore,
the longitudinal resistance in this theory will behave as $R_{xx}\sim T^{-6/5}$,
precisely as in the Abelian hierarchy state. However, shot noise experiments
\cite{Picciotto97,Saminadayar97,Dolev:cm08}
can detect the charge of the tunneling quasiparticles.
In the Abelian hierarchy state, the current is carried by
charge $2e/5$ quasiparticles at the lowest temperatures, where the
most relevant operator (in the RG sense) will dominate.
In the non-Abelian hierarchy state, charge $e/5$ quasiparticle tunneling
is the most relevant operator. In the $k=3$ $\overline{\rm RR}$ state,
charge $e/5$ and charge $2e/5$ quasiparticle tunneling are equally
relevant, but the bare tunneling matrix element for charge $e/5$
quasiparticles is presumably larger than for
charge $2e/5$ quasiparticles ($\Gamma_{2/5}\sim (\Gamma_{1/5})^2$),
so tunneling will be dominated by the former. In summary, we expect
shot noise experiments in either of the non-Abelian states to
result in a charge of $e/5$, as compared to charge $2e/5$ in the
Abelian state. The two non-Abelian states can be distinguished
from each other by the power-laws with which $R_{xx}$ depends
on $T$ or $I_{\rm tun}$ on $V$ for $V>T$ in the limit of weak tunneling.
In the opposite limit of strong tunneling, the droplet effectively breaks in two and
all that remains is the weak tunneling of electrons between the two droplets.
In this case, $G\sim T^{4{\Delta_e}-2}$; in both the Abelian and non-Abelian hierarchy
states, ${\Delta_e}=3/2$ while in the $k=3$ $\overline{\rm RR}$ state, ${\Delta_e}=2$.

We acknowledge helpful discussions with Jim Eisenstein, Lukasz Fidkowski, Gil Refael,
and especially Eddy Ardonne. GAF was supported by the Lee A. DuBridge Foundation
and the NSF grant PHY05-51164.

\vskip -0.6 cm


\end{document}